\documentclass[apjl]{emulateapj}

\usepackage{epstopdf}
\usepackage{natbib}
\usepackage{amsmath}
\usepackage{enumerate}
\usepackage{hyperref}

\usepackage{cancel}

\usepackage{graphicx}
\usepackage{color}
\usepackage{mathtools}

\def\simpropto{\lower.2ex\hbox{$\; \buildrel \propto \over \sim \;$}}
\def\ltsim{\lower.5ex\hbox{$\; \buildrel < \over \sim \;$}}
\def\gtsim{\lower.5ex\hbox{$\; \buildrel > \over \sim \;$}}

\usepackage{color}

\def\bea{\begin{eqnarray}}
\def\eea{\end{eqnarray}}
\def\simpropto{\lower.2ex\hbox{$\; \buildrel \propto \over \sim \;$}}
\def\ltsim{\lower.5ex\hbox{$\; \buildrel < \over \sim \;$}}
\def\gtsim{\lower.5ex\hbox{$\; \buildrel > \over \sim \;$}}
\begin{document}
\def\mnras{Mon. Not. Roy. Astr. Soc.}
\def\ap{ApJ}
\def\apjl{ApJL}
\def\jcap{JCAP}

\title{Intermediate mass black holes and dark matter at the Galactic center}

\author{Thomas Lacroix}
\affiliation{Laboratoire Univers \& Particules de Montpellier (LUPM),
  CNRS \& Universit\'e de Montpellier (UMR-5299),
  Place Eug\`ene Bataillon,
  F-34095 Montpellier Cedex 05, France}
\author{Joseph Silk}
\affiliation{Institut d'Astrophysique de Paris, UMR 7095, CNRS \& UPMC, Sorbonne Universit\'{e}s, 98 bis bd Arago, F-75014 Paris, France}
\affiliation{Department of Physics and Astronomy, The Johns Hopkins University,
 Baltimore, MD 21218, USA}
\affiliation{Beecroft Institute of Particle Astrophysics and Cosmology, Department of Physics, University of Oxford, Oxford OX1 3RH, UK}
\today

\begin{abstract}
Could there be a large population of intermediate-mass black holes (IMBHs) formed in the early universe?  Whether primordial or formed  in Population III, these are likely to be very subdominant compared to the dark matter density, but could seed early dwarf galaxy/globular cluster and supermassive black hole formation. Via survival of dark matter density spikes, we show here that  a centrally concentrated relic population of IMBHs, along with ambient dark matter, could account for the \textit{Fermi} gamma-ray ``excess" in the Galactic center because of dark matter particle annihilations. 
\end{abstract}

\keywords{dark matter -- Galaxy: center -- stars: black holes}


\section {Introduction}

The diffuse \textit{Fermi}-LAT ``excess" (e.g., \citealp{goodenough09,ajello2016}) or residual emission from the Galactic center (GC) remains the strongest evidence for dark matter (DM) that consists of weakly interacting massive particles (WIMPs). The hypothesis of weakly annihilating supersymmetry-motivated WIMPs is well supported by particle physics arguments, despite the lack of success in finding any evidence for supersymmetry. The morphology, density profile, and spectrum of the \textit{Fermi} excess collectively support a DM interpretation.

However, recent analyses of the $\gamma$-ray statistical fluctuations demonstrate that any diffuse DM contribution must be subdominant. One needs to include a discrete source component  of $\sim 1000$ subthreshold sources to account for the results of a fluctuation analysis \citep{bartels16, lee16}. The leading candidate for such a component is an old population of millisecond pulsars (MSPs), as detected in massive globular clusters. If such a component can account not only for the fluctuations but the other observed  characteristics of the diffuse excess, notably morphology, density profile, and spectrum, then the motivation for any DM component is largely removed.

Any such interpretation may be premature, if only because the population of weak MSPs is not well known. However, it is difficult for a 
DM self-annihilation model to be sufficiently clumpy. One such attempt involves strongly interacting DM that can in principle form a population of dark compact objects \citep{agrawal17}.

Here, we propose an arguably more compelling clumpy DM model, in which we combine a population of IMBHs, originating either as Population III (Pop III) or primordial black holes (PBHs), with the standard WIMP DM model. The latter  features the usual weakly interacting self-annihilating DM particles that  have successfully reproduced the  spectrum, profile, and morphology of the \textit{Fermi} GC excess, but at the same time are challenging direct detection experiments (e.g., \citealp{aprile17}) and indirect detection of $\gamma$-rays from nearby dwarfs \citep{ackermann15}.

Our motivation is that DM WIMPs are plausible from the perspective of particle physics, and a highly subdominant population of IMBHs is equally plausible from astrophysics.   Single field inflation models provide a strong case for PBH formation, when at the end of slow roll, there is a generic phase prior to reheating when small scale but high amplitude fluctuations are generated. These plausibly form rare PBHs \citep{carr74}. The predicted mass spectrum extends from the Hawking evaporation limit, about $10^{16}\, \rm g$ to $10^5\, M_\odot$ or larger. PBHs as a DM contributor are compelling from astrophysical arguments, as one just needs Einstein gravity and essentially standard cosmology, without need for new particle physics. The primordial fluctuation mass range generating PBHs as a subdominant DM contributor is  unconstrained by cosmic microwave background (CMB) or large-scale structure observations, which only constrain galactic scales or larger. Constraints on relevant scales are eventually possible  from limits on CMB spectral distortions and stochastic gravity wave backgrounds.

The astrophysical motivation for an early universe population of IMBHs formed during the Pop III epoch is equally compelling. Recent observations favor a significant, although highly subdominant, population of massive central BHs in dwarf galaxies. These are observed as active galactic nuclei (AGNs), with X-ray, optical, and infrared signatures, and an occupation number of order 1\% \citep{baldassare}. Hence the IMBH occupation number, allowing for a duty cycle, must be significantly higher, by at least an order of magnitude. Simulations of formation of SMBH by mergers of IMBHs suggest that $\Omega_{\mathrm{IMBH}} \sim 10^{-3} \Omega_{\mathrm{baryon}}$ \citep{islam04,rashkov14}.

The IMBHs may be PBHs. This is an interesting, although not obligatory, option. Dwarf galaxy IMBHs  need not be PBHs, although  it is difficult to populate dwarfs, other than first-generation systems, with central massive BHs. Formation of massive IMBHs by merging of smaller BHs generates recoils. Either way, we are likely to have a highly subdominant population of IMBHs in massive galaxies. For a Pop III origin, a simple estimate of the mass fraction of first-generation dwarfs is of order 0.001, based on the Pop III metallicity threshold. If a significant fraction of these undergo direct collapse (options are suppression of key coolant $\mathrm{H}_2$ by UV from neighboring Pop III embryonic dwarfs, \citealp{regan2017}, or supercritical Pop III BH accretion; \citealp{lupi2016,begelman2017}), one coincidentally would arrive at the IMBH mass fraction required to account for the dwarf galaxy IMBHs as observed. 

The idea that PBHs could be 1\% or more of DM has been revived by the aLIGO detection of four confirmed binary merging BHs of mass 10--30 $M_\odot$, although the predicted mass fraction of PBHs required is model-dependent, ranging from $\sim$100\% \citep{bird16} to $\sim$1\% \citep{sasaki16, alihaimoud17}. Observations, most notably from gravitational microlensing \citep{alcock01}, dwarf galaxy heating \citep{brandt16} and CMB distortions \citep{alihaimoud17a}, favor the lower range.  Future aLIGO observations, combined with Virgo and other detectors, should be able to distinguish PBHs from the more conventional astrophysical explanation via the observed BH mass distribution.  However, for the purposes of this Letter, a mass fraction of only 0.1\% in  intermediate-mass PBHs would be required, if indeed we have greatly overestimated the Pop III contribution.

More generally, a case can be made for massive BHs in all dwarfs that subsequently merge hierarchically as in the usual cold dark matter (CDM) model to form more massive galaxies.  Firstly, mergers of IMBHs form a natural,  but admittedly inefficient, path to forming supermassive black holes (SMBHs), given the ubiquity of mergers in structure formation.  Second, the most massive SMBHs, if formed by accretion at or below the Eddington limit, require seed IMBHs  \citep{habouzit16}. Third, many, if not all, of the problems in dwarf galaxy formation scenarios, including abundance, cusp/core controversy, too-big-to-fail, and baryon fraction, can potentially be resolved by the additional degrees of freedom provided by AGN feedback \citep{silk17}, without recourse to exotic DM models.

Recent observations point to IMBHs in massive globular clusters \citep{kiziltan17}, which could provide additional indicators of their presence in low-mass systems. If indeed all merging substructures, to be envisaged for example as protoglobular clusters in typical bulge formation models or more generally,  protodwarf galaxies, contain {IMBHs}, then not only does this provide a natural pathway for forming the central nuclear star cluster (NSC) and SMBH along with  the bulge and stellar halo  in the Milky Way Galaxy (MWG; \citealp{antonini15}), but also a robust prediction: there should be a large population of massive BHs that failed to merge \citep{islam04, rashkov14}.

For IMBHs formed in the early universe, whatever their origin, DM density spikes are inevitable, with a profile $\rho(r) \propto r^{-\gamma_{\mathrm{sp}}}$ within the BH influence radius, where $\gamma_{\mathrm{sp}} > 3/2$. The limiting value is $\gamma_{\mathrm{sp}} = 9/4$ for an isothermal DM core, up to a plateau determined by the annihilation rate \citep{gondolo99}, but spikes around primordial IMBHs can be even steeper \citep{Eroshenko16}. Correction for the effect of mergers on the density profile \citep{Merritt2002} remains an uncertainty, although spikes can regenerate via accretion. Moreover, flattening within the BH influence radius occurs to $\gamma_{\mathrm{sp}} \approx 3/2$ due to stellar heating \citep{gnedin04} in the case of the SMBH at the GC, but it is not clear how this affects IMBHs or, in particular, the near-horizon density. One may wonder about the impact of a putative spike around the SMBH at the GC on the $\gamma$-ray emission. In this model, the SMBH formed by mergers. The IMBHs are the merged relics. Hence, the SMBH spike would have been destroyed or softened via dynamical heating from the mergers (like in \citealp{Merritt2002}). The central spike may also have been dynamically heated in the NSC \citep{gnedin04}. Finally, the radial extension of the SMBH spike may also be sufficiently small for the associated $\gamma$-ray emission to be subdominant. In the end, we may have a significant contribution to the $\gamma$-ray emission from the mini-spikes and a subdominant one from the central SMBH spike.

The beauty of self-annihilating DM density spikes is that one can probe very low cross-sections, leading to  unique constraints, as found for the case of M87 \citep{lacroix15}. Our model is largely motivated  by dwarf galaxy observations that  show a  possibly high occupation number of massive BHs in dwarfs. One attractive feature of IMBHs is that, being in place before MWG-like  galaxies, they can act, especially if PBHs, as seeds of dwarfs as well as of SMBHs and even of more massive systems \citep{clesse15}. PBHs may form via primordial non-Gaussian features in the initial fluctuation spectrum \citep{young16}, with possible implications for the stochastic gravitational-wave background \citep{nakama17}. Early formation of dwarfs has previously been invoked to address issues of reionization of the universe  via primordial non-Gaussianities \citep{habouzit14}, and the relative roles of AGNs and massive stars in reionization are in principle distinguishable \citep{cohen}. Such dwarfs most likely contain IMBHs if we accept the growing body of astrophysical evidence on nearby dwarfs. 

Merging dwarf galaxies would generate a massive BH distribution that is steeper than that of the initial CDM-like profile, due to dynamical friction, and might plausibly approximate that of the stellar bulge, whose radial profile is $\propto r^{-2}$,{ which turns out to be possibly more consistent with the profile of the residual $\gamma$-ray emission}.  An unavoidable consequence would also be the formation of an NSC by the most tightly bound stars \citep{antonini15}. These are likely to have been more enriched than typical dwarf, hence bulge, stars, leading to a possible chemical discriminant. 

\section{Gamma rays from DM spikes around IMBHs}
\label{normalization}
We give simple numerical estimates that illustrate how the IMBH-spike scenario can readily account for the \textit{Fermi} ``excess" for very small annihilation cross-sections. Following  \citet{agrawal17}, we  estimate the DM parameters that can reproduce the flux of one point source from the analysis of \citet{lee16}. About $10^{3}$ such point sources are then needed to contribute of the order of the residual $\gamma$-ray emission.
The integrated photon flux for a spike around an individual IMBH---referred to as a mini-spike in the following---between $E_{\gamma,\mathrm{min}} = 1.893\, \mathrm{GeV}$ and $E_{\gamma,\mathrm{max}} = 11.943\, \mathrm{GeV}$ \citep{lee16} is given as usual by 
\begin{equation}
\Phi_{\mathrm{sp}} = \dfrac{\left\langle \sigma v \right\rangle}{2 m_{\mathrm{DM}}^{2} d^{2}} \left( \int_{E_{\gamma,\mathrm{min}}}^{E_{\gamma,\mathrm{max}}} \! \dfrac{\mathrm{d}N}{\mathrm{d}E_{\gamma}} \, \mathrm{d}E_{\gamma}\right) \left( \int_{0}^{R_{\mathrm{sp}}} \! r^{2} \rho^{2}(r) \, \mathrm{d}r \right),
\end{equation}
where $\left\langle \sigma v \right\rangle$ is the velocity-averaged annihilation cross-section, $m_{\mathrm{DM}}$ the mass of the DM candidate, $\mathrm{d}N/\mathrm{d}E_{\gamma}$ the $\gamma$-ray spectrum per annihilation---taken from \citet{cirelli2011}, and $d \approx 8.32\ \rm kpc$ \citep{gillessen2017}, the distance between Earth and the IMBH. Our benchmark scenario is a DM candidate with $m_{\mathrm{DM}} = 30\ \rm GeV$ annihilating into $b\bar{b}$, compatible with the spectral properties of the GC residual $\gamma$-ray emission. The DM profile in the mini-spike is defined as follows:\footnote{We use a simplified expression based on \citet{gondolo99}. The inner boundary of the spike is related to capture of DM particles by the BH and is given by $2R_{\mathrm{S}}$ for a non-rotating BH \citep{sadeghian13}. The precise value of this inner radius has no impact on our calculations. The DM profile outside the spike is not relevant since the dominant part of the annihilation flux comes from the inner region of the spike.} 
\begin{equation}
\rho(r) = 
\begin{cases}
0 & r \leqslant 2 R_{\mathrm{S}} \\
\rho_{\mathrm{sat}} & 2 R_{\mathrm{S}} < r \leqslant R_{\mathrm{sat}} \\
\rho_{0} \left( \dfrac{r}{R_{\mathrm{sp}}} \right)^{-\gamma_{\mathrm{sp}}} & R_{\mathrm{sat}} < r \leqslant R_{\mathrm{sp}}
\end{cases},
\end{equation}
where the saturation density is given by $\rho_{\mathrm{sat}} = m_{\mathrm{DM}}/(\left\langle \sigma v \right\rangle t_{\mathrm{BH}})$ with $t_{\mathrm{BH}}$ the  BH age, and $R_{\mathrm{sat}} = R_{\mathrm{sp}} (\rho_{\mathrm{sat}}/\rho_{0})^{-1/\gamma_{\mathrm{sp}}}$ by continuity. The radial extension of the spike $R_{\mathrm{sp}}$ is of the order of the BH influence radius, $G M_{\mathrm{BH}}/\sigma_{*}^{2}$ \citep{peebles72}. The extended $M_{\mathrm{BH}}$-$\sigma_{*}$ relation for IMBHs \citep{tremaine2002} gives an estimated value of $\sigma_{*} \approx 10\ \rm km\ s^{-1}$, and $R_{\mathrm{sp}} \approx 0.043\, \rm pc$. Then, $\rho_{0} \approx (3-\gamma_{\mathrm{sp}}) M_{\mathrm{sp}}/(4 \pi R_{\mathrm{sp}}^{3})$ by the requiring the mass inside the spike $M_{\mathrm{sp}}$ be of the order of the BH mass, with $M_{\mathrm{sp}} \approx M_{\mathrm{BH}} \approx 10^{2}$--$10^{3}\  M_{\odot}$. For $\gamma_{\mathrm{sp}} > 3/2$, the integrated flux for a single mini-spike reads\footnote{For $\gamma_{\mathrm{sp}} = 3/2$, the integrated flux for a single mini-spike is given by
\begin{align}
\Phi_{\mathrm{sp}} \approx \ & 7 \times 10^{-12} \, \mathrm{cm^{-2} \, s^{-1}} \ln \left( 6 \times 10^{6} \left( \dfrac{M_{\mathrm{sp}}}{10^{3}\, \mathrm{M_{\odot}}} \right)^{-1}  \left( \dfrac{R_{\mathrm{sp}}}{0.043\, \mathrm{pc}} \right)^{3} \right. \nonumber \\
& \times \left. \left( \dfrac{m_{\mathrm{DM}}}{30\, \mathrm{GeV}} \right) \left( \dfrac{\left\langle \sigma v \right\rangle}{3 \times 10^{-31} \, \mathrm{cm^{3}\, s^{-1}}} \right)^{-1} \left( \dfrac{t_{\mathrm{BH}}}{10^{10}\, \mathrm{yr}} \right)^{-1}  \right) \nonumber \\
& \times \left( \dfrac{d}{8.32\, \mathrm{kpc}} \right)^{-2} \left( \dfrac{M_{\mathrm{sp}}}{10^{3}\, \mathrm{M_{\odot}}} \right)^{2} \left( \dfrac{R_{\mathrm{sp}}}{0.043\, \mathrm{pc}} \right)^{-3} \nonumber \\
& \times \left( \dfrac{m_{\mathrm{DM}}}{30\, \mathrm{GeV}} \right)^{-2} \left( \dfrac{\left\langle \sigma v \right\rangle}{3 \times 10^{-31} \, \mathrm{cm^{3}\, s^{-1}}} \right) \left( \dfrac{N_{\gamma}^{(\mathrm{tot})}}{5.7} \right). \nonumber
\end{align}}
\begin{align}
\label{spike_flux}
\Phi_{\mathrm{sp}} =\ & \dfrac{\gamma_{\mathrm{sp}}}{3(2 \gamma_{\mathrm{sp}}-3)} \left( \dfrac{3-\gamma_{\mathrm{sp}}}{4\pi} \right)^{\frac{3}{\gamma_{\mathrm{sp}}}} \dfrac{1}{d^{2}} M_{\mathrm{sp}}^{\frac{3}{\gamma_{\mathrm{sp}}}} \nonumber \\
& \times R_{\mathrm{sp}}^{3\left( 1-\frac{3}{\gamma_{\mathrm{sp}}}\right) } t_{\mathrm{BH}}^{\frac{3}{\gamma_{\mathrm{sp}}}-2} m_{\mathrm{DM}}^{-\frac{3}{\gamma_{\mathrm{sp}}}} \left\langle \sigma v \right\rangle^{\frac{3}{\gamma_{\mathrm{sp}}}-1} N_{\gamma}^{(\mathrm{tot})},
\end{align}
where 
$N_{\gamma}^{(\mathrm{tot})} = \int_{E_{\gamma,\mathrm{min}}}^{E_{\gamma,\mathrm{max}}} \! (\mathrm{d}N/\mathrm{d}E_{\gamma}) \, \mathrm{d}E_{\gamma}$. More specifically, for $\gamma_{\mathrm{sp}} = 9/4$, the flux from a mini-spike is
\begin{align}
\Phi_{\mathrm{sp}} \approx & \ 1 \times 10^{-10}\, \mathrm{cm^{-2}\,s^{-1}} \left( \dfrac{d}{8.32\, \mathrm{kpc}} \right)^{-2} \left( \dfrac{M_{\mathrm{sp}}}{10^{3}\, \mathrm{M_{\odot}}}\right) ^{4/3} \nonumber \\
& \times \left( \dfrac{R_{\mathrm{sp}}}{0.043\, \mathrm{pc}} \right)^{-1} \left( \dfrac{m_{\mathrm{DM}}}{30\, \mathrm{GeV}} \right)^{-4/3} \nonumber \\
& \times \left( \dfrac{\left\langle \sigma v \right\rangle}{2 \times 10^{-40} \, \mathrm{cm^{3}\, s^{-1}}} \right)^{1/3} \left( \dfrac{t_{\mathrm{BH}}}{10^{10}\, \mathrm{yr}} \right)^{-2/3} \left( \dfrac{N_{\gamma}^{(\mathrm{tot})}}{5.7} \right).
\end{align}
Let us assume that the entire point-source contribution to the \textit{Fermi} excess fluctuations is given by the IMBH spikes. The resulting limits on the annihilation cross-section are as follows. The upper limit on $\left\langle \sigma v \right\rangle$ is extremely small ($\sim 10^{-40}\, \mathrm{cm^{3}\, s^{-1}}$) for a steep mini-spike with $\gamma_{\mathrm{sp}} = 9/4$ and $M_{\mathrm{BH}} = 10^{3}\, M_{\odot}$. This is related to the very weak dependence of $\Phi_{\mathrm{sp}}$ on the cross-section. For a relaxed spike with $\gamma_{\mathrm{sp}} = 3/2$, the upper limit on the cross-section is of the order of $10^{-31} \, \mathrm{cm^{3}\, s^{-1}}$ for a population of $10^{3}\, M_{\odot}$ IMBHs. For $M_{\mathrm{BH}} = 10^{2}\, M_{\odot}$, the best-fit cross-sections become $2 \times 10^{-36}\, \mathrm{cm^{3}\, s^{-1}}$ for $\gamma_{\mathrm{sp}} = 9/4$ and $3 \times 10^{-29}\, \mathrm{cm^{3}\, s^{-1}}$ for $\gamma_{\mathrm{sp}} = 3/2$.

\section{Global signal and spatial morphology}
We now consider a distribution of IMBHs that collect in the inner galaxy. Most of them are failed mergers, as mentioned above, with a total mass amounting to of the order of the mass of the central SMBH, $4 \times 10^6\, M_\odot$. First, we note that to compute the radial profile of $\gamma$-rays, we need to convolve the radial distribution of the IMBHs with the radial dependence of the mini-spike flux.

The DM interpretation of the \textit{Fermi} excess works for the morphology because it naturally gives the $\gamma$-ray profile as roughly the square of the NFW profile, or $r^{-2}$. The present model needs to address this point. However, the case for the DM interpretation may not be that strong. Firstly, the MWG may have a DM core \citep{Portail17}. Second, the \textit{Fermi} excess can be fit, according to a reanalysis, by a stellar mass (bulge)-related profile \citep{bartels17}.  

The point sources (IMBHs) have a $r^{-3/2}$ density profile toward the GC. This  is a dynamically relaxed profile that follows the Bahcall--Wolf solution for a stellar cusp \citep{BW76}. This might not match the observed profile if the mini-spike masses and luminosities are independent of radius. In fact,
there will be mass segregation, the more massive IMBHs falling in closer to the GC but stalling at/near the final parsec. IMBHs are point masses, and too dense to be tidally disrupted. Let us estimate the radial dependence of the mini-spike luminosity. The radial flux profile for a set of mini-spikes around BHs is $\Phi_{r} \propto \Phi_{\mathrm{sp}} r^{-3/2}$. From Eq.~(\ref{spike_flux}), the flux for an individual mini-spike is $\Phi_{\rm sp}  \propto M_{\mathrm{BH}}^{3-6/\gamma_{\rm sp}}$, where $\gamma_{\mathrm{sp}}= (9-2\gamma)/(4-\gamma)$. Here, we assume $M_{\mathrm{sp}} = M_{\mathrm{BH}}$  and $R_{\mathrm{sp}} =GM_{\mathrm{BH}}/\sigma_{*}^2$, with $\sigma_{*} \propto M_{\mathrm{BH}}^{1/4}$ \citep{tremaine2002} so that $R_{\mathrm{sp}} \propto M_{\mathrm{BH}}^{1/2}$. Hence, $\Phi_{\mathrm{sp}}  \propto M_{\mathrm{BH}}^{(3/2)(1-1/\gamma_{\mathrm{sp}})}$. In addition, $M_{\mathrm{BH}}$ increases as $r$ decreases because of mass segregation by settling. The two-body relaxation time-scale is $\propto t_{\mathrm{r}} \propto 1/M_{\mathrm{BH}}$, as is the dynamical friction time that is $\sim M_{\mathrm{SMBH}}/M_{\mathrm{BH}}$ orbital times. One needs a simple diffusion model to go further, but a rough guess using adiabatic invariants might be $r v M_{\mathrm{BH}} = \mathrm{const}$ (conservation of angular momentum), so that $M_{\mathrm{BH}} \propto r^{-1/2}$. Hence, the radial flux profile is 
\begin{equation}
\Phi_{r} \propto r^{-\frac{9}{4} \left( 1 - \frac{1}{3\gamma_{\mathrm{sp}}} \right) }.
\end{equation}
Generally, $\gamma_{\mathrm{sp}}= (9-2\gamma)/(4-\gamma)$, so that for a core, $\gamma=0$ and $\gamma_{\mathrm{sp}}=9/4$, while for $\gamma = 1$, $\gamma_{\mathrm{sp}}=7/3$ and for $\gamma=3/2$, $\gamma_{\mathrm{sp}}=12/5$. Hence, for adiabatic mini-spikes, the radial profile is $\Phi_{r}^{(\gamma=0)} \propto r^{-23/12}$, $\Phi_{r}^{(\gamma=1)} \propto r^{-27/14}$ and $\Phi_{r}^{(\gamma=3/2)} \propto r^{-31/16}$. Therefore, $\Phi_{r} \propto r^{-2}$ for IMBH mini-spikes, always approximating the observed $\gamma$-ray profile independently of the DM halo profile.

\begin{figure}[h!]
\centering 
\includegraphics[width=\linewidth]{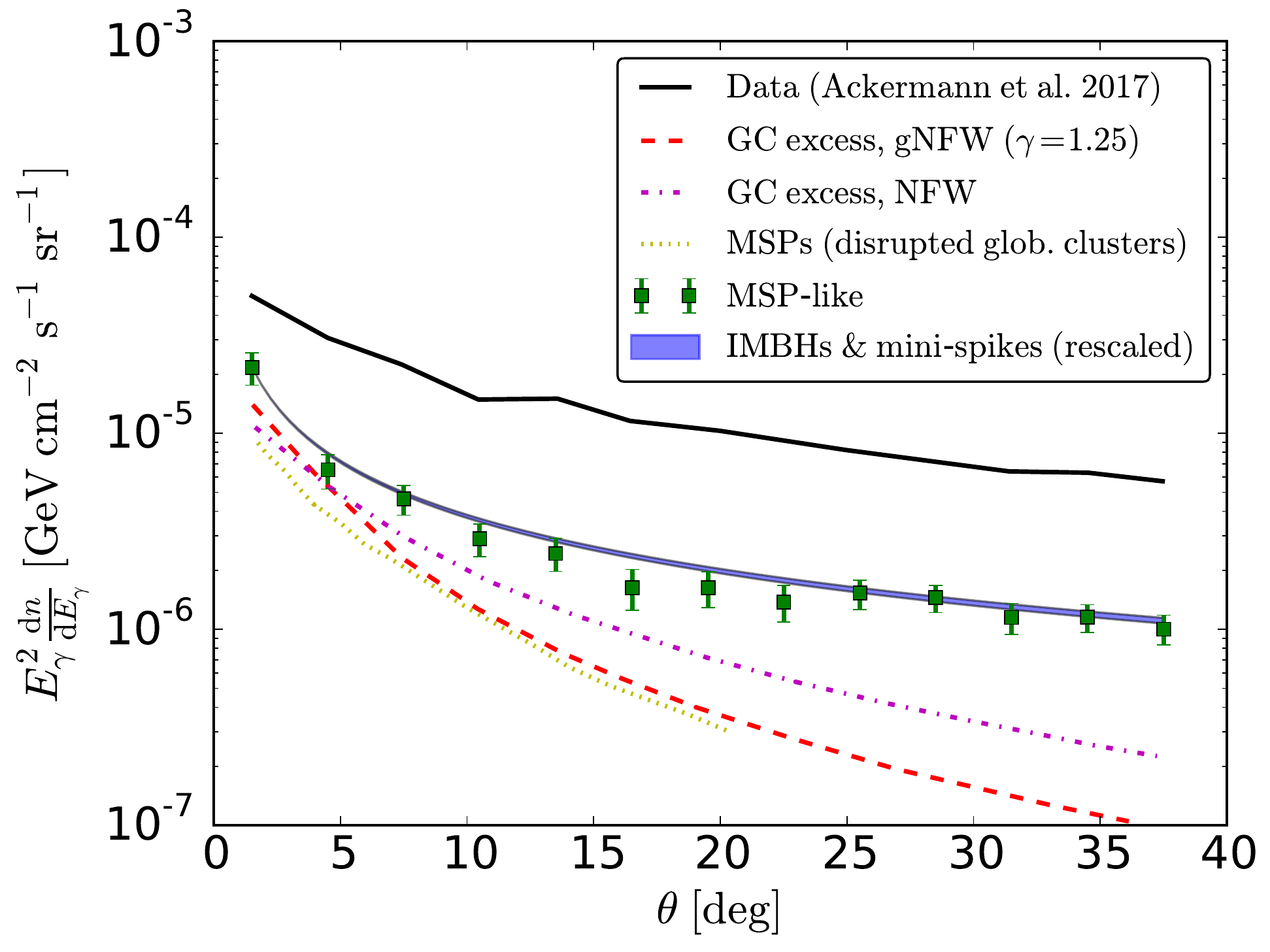} 
\caption{\label{profile}Angular profiles for the total $\gamma$-ray emission at 2 GeV with bright point sources masked (black solid line; \citealp{ackermann17}) and for various components of the $\gamma$-ray emission. Green squares: MSP-like component extracted from the data \citep{ackermann17}. Red dashed line: GeV excess in the sample model from \citet{ackermann17} corresponding to a generalized NFW profile template with slope $\gamma = 1.25$. Magenta dotted-dashed line: GeV excess in the sample model but for a regular NFW profile ($\gamma = 1$). Yellow line: prediction for MSPs in the bulge of the Milky Way from disrupted globular clusters \citep{brandt15}. Our IMBH--mini-spike model is depicted by the blue shaded area for benchmark slopes discussed in the text. Here, we are mostly interested in illustrating the spatial morphology of the signal in our model, so we arbitrarily rescaled the angular IMBH--mini-spike profile at the level of the first MSP-like point.}
\end{figure}

These predictions are illustrated in Fig.~\ref{profile}, which shows the angular profile of the total GC $\gamma$-ray data at 2 GeV with bright point sources masked \citep{ackermann17}, along with the profiles of various components of the $\gamma$-ray emission. Our model typically gives an angular profile that is consistent with expectations from bulge sources like MSPs. 

The situation is complicated by the fact that the DM spikes may be heated---for relaxed mini-spikes $\Phi_{r} \propto r^{-1.75}$---and partially stripped as the IMBHs fall into the GC region, although tidal disruption of PBH clusters and dynamical friction may in turn steepen the IMBH profile, as discussed in \citet{fragione17} for MSPs. Regardless, it seems plausible, pending detailed simulations, that our model gives a good approximation to the \textit{Fermi} $\gamma$-ray excess profile.

\section{Discussion}
One  attractive model  for the LIGO events argues that hard massive BH binaries form in dense stellar clusters. This scenario has one advantage over rivals: it was proposed before the aLIGO detection \citep{bae14} to give acceptable rates and masses. Protoglobular clusters are likely pregalactic sites
 and are dispersed as substructure disrupts when the  bulge formed. Stellar cluster-enhanced  formation of massive BH binaries quantitatively accounts for the observed LIGO rates, when integrated  out to several hundred Mpc \citep{park17}.  Such massive BHs may have formed prolifically at high redshift, when there was most likely a top-heavy initial mass function, providing  a possible pathway to forming IMBHs. In the PBH case, one appeals to BH binary formation by early capture in the first bound DM substructures  at the onset of matter domination \citep{sasaki16}. Some subsets of these (one needs of the order of 10\%) might have merged to form IMBHs.

We expect that massive binaries should be enhanced in number near the GC where the most massive protoglobulars dispersed to form the central NSC. These would generate MSPs as well as BH binaries. Hence, these two populations should track each other. Neither would have a significant disk component. Another consequence would be an enhanced rate of BH mergers in galactic nuclei that might be detectable by LIGO \citep{nishikawa2017}. These LIGO events  occur  within the IMBH sphere of influence. This could lead to enhanced drag and affect the gravitational-wave signal phase evolution. This could potentially be seen as a cumulative phase shift by LISA over many cycles \citep{yue2017}.

We showed that mini-spikes around a population of  hundreds or thousands  IMBHs can significantly contribute to the GC emission and can readily account for both the normalization and spatial morphology of the $\gamma$-ray excess for very small annihilation cross-sections. The expected morphology of the predicted excess  does not necessarily follow the standard DM halo profile, for instance, it can effectively trace the Galactic bulge due to mass segregation and the dependence of mini-spike luminosities on BH mass. This circumvents the issue raised by the observation of an excess of $\gamma$-rays in control regions in the disk where no significant contribution from DM is expected \citep{ackermann17}. IMBHs would appear naturally in central regions due to three-body encounters and ejections. This distinctive morphology also allows the model to evade the constraints of \citet{clark16} that ruled out a DM interpretation of the excess in terms of ultra-compact mini-halos. We note that the constraints of \citet{clark16} do not account for more recent studies of the GC emission that revealed a more complex spatial morphology \citep{ackermann17,bartels17}. Finally, we expect the central massive BHs seen in nearby galactic centers, if indeed formed in the early universe, to have DM spikes, and hence to be \textit{Fermi} $\gamma$-ray sources. 47 Tuc  is a possible example \citep{abdo09}, although one cannot easily distinguish a possible $\gamma$-ray point source from the expected  population of MSPs. Future observations may help us elucidate this point.

\section*{Acknowledgments}
T.L. receives financial support from CNRS-IN2P3. T.L. also acknowledges support from the European Union’s Horizon 2020 research and innovation program under the Marie Sk\l{}odowska-Curie grant agreement Nos. 690575 and 674896; beside recurrent institutional funding by CNRS-IN2P3 and the University of Montpellier.  The work of J.S. has been supported in part by European Research Council (ERC) Project No. 267117 (DARK) hosted by Universit\'{e} Pierre \& Marie Curie -- Paris VI, Sorbonne Universit\'{e}s.

%

\end{document}